# A Se vacancy induced localized Raman mode in two-dimensional MoSe$_2$ grown by CVD

Shudong Zhao[1]‡, Meilin Lu[1]‡, ShaSha Xue[1], Lin Yan[1], Peng Miao[2], Yan Hang[3], Xianjie Wang[1], Zhiguo Liu[1], Yi Wang[4], Lei Tao[5], Yu Sui[1] and Yang Wang[4]

[1] Department of Physics, Harbin Institute of Technology, Harbin 150001, People's Republic of China.

[2] School of Chemistry and Chemical Engineering, Harbin Institute of Technology, Harbin 150001, People's Republic of China.

[3] School of Materials Science and Engineering, Harbin Institute of Technology, Harbin 150001, People's Republic of China

[4] Academy of Fundamental and Interdisciplinary Sciences, Harbin Institute of Technology, Harbin 150001, People's Republic of China.

[5] Laboratory for Space Environment and Physical Sciences, Harbin Institute of Technology, Harbin 150001, People's Republic of China

‡ Shudong Zhao and Meilin Lu contributed equally to this work

E-mail: taolei@hit.edu.cn (Lei Tao); suiyu@hit.edu.cn (Yu Sui); yangwang@hit.edu.cn (Yang Wang)

**Abstract**

Defects play a significant role in optical properties of semiconducting two-dimensional transition metal dichalcogenides (TMDCs). In ultra-thin MoSe$_2$, a remarkable feature at ~250 cm$^{-1}$ in Raman spectra is ascribed to be a defect-related mode. Recent attempts failed to explain the origin of this peak, leaving it being a mystery. Here in this work, we demonstrate that this peak is a Se vacancy induced defect mode. Heat effect and hydrogen etching are two main factors to introduce Se vacancies in CVD process of growing MoSe$_2$. A phonon confinement model can well explain the behaviors of intrinsic Raman modes. Density functional theory (DFT) calculation reveals that single Se vacancy (V$_{Se}$) is responsible for the appearance of Raman peak at ~250 cm$^{-1}$ and this mode is an A$_{1g}$-like localized mode which is also confirmed by polarized Raman scattering experiment. The relative strength of this mode can be a characterization of the quality of 2D MoSe$_2$. This work may offer a simple method to tailor chalcogenide vacancies in 2D TMDCs and provide a way to study their vibrational properties.

# 1. Introduction

Two-dimensional (2D) transition metal dichalcogenides (TMDCs), in particular $MoS_2$, $MoSe_2$, $WS_2$ and $WSe_2$, have shown novel properties in many aspects, such as direct-bandgap nature at monolayer scale[1, 2], large exciton binding energy[3, 4], large carrier mobility [5, 6], valley polarization[7], room-temperature ferromagnetism[8] and so on. These properties make 2D TMDCs a kind of very promising material for applications in electronic, optoelectronic, valley electronics or spintronic devices. Although great achievements have been made in field effect transistors[6, 9-11], light-emitting diodes[12], CMOS or large-scale integrated circuits[13, 14], some fundamental physics of these materials are still less known, especially in structure of defects and roles of defects in electrical, magnetic or optical properties of 2D TMDCs. Recently, structure of defects in TMDCs have been characterized by STEM and STM measurement. However, these two methods are expensive and have strict requirements on sample, which may even cause damages of the ultra-thin TMDCs in process of sample preparation or electron irradiation[15, 16].

Raman spectrum has been proved to be a fast, convenient and non-destructive method to characterize the defects in 2D TMDCs. Previously in $MoS_2$ and $WS_2$, defects has been proved to lead to shifting of characteristic peaks and appearance of new defect-related peaks in Raman spectra[17-19]. As to $MoSe_2$, except two main features of Raman scattering, peak ~ 240 $cm^{-1}$ for $A_{1g}$ mode and peak ~ 288 $cm^{-1}$ for $E^1_{2g}$ mode[20], a remarkably mysterious peak around 250 $cm^{-1}$ may or may not show in Raman spectra of the as-prepared samples, obtained by either exfoliation or CVD methods as far as our best survey is concerned[21-36]. Moreover, its relative intensity varies in different works, so it is reasonable to draw a clue that this unknown peak is related to the quality of the $MoSe_2$ nanosheets synthesized by different methods in different research groups. Nevertheless, the assignment of this peak is still controversial.

Recently, in Ismail's work the Raman peak at 250 $cm^{-1}$ was assigned to the double resonance of ZA mode at M point Brillouin zone that appears in high-quality thin $MoSe_2$ films using $MoO_2$ powders instead of $MoO_3$ as precursor[36]. But in general, the appearance of a new peak in Raman spectra is usually associated with some kind of lattice disorder. As reported in Masoud's work, the peak at 250 $cm^{-1}$ is associated with Se vacancies which are introduced by laser heating in their laser-based synthesis approach. STEM results confirmed two structures of Se vacancies, i.e. single Se vacancy ($V_{Se}$) and dual Se column vacancy ($V_{2Se}$). However, their density functional theory (DFT) calculation results failed to assign this peak using the supercells with different $V_{Se}$ concentrations[37].

Here, in order to make sure what cause Se vacancies in CVD prepared 2D $MoSe_2$ and what structure of Se vacancy is and how Se vacancy induces this mysterious peak at ~250 $cm^{-1}$ in Raman spectrum, we adopted CVD method using $MoO_3$ and Se powders as precursors like most other works did to synthesize $MoSe_2$ nanosheets[21, 29]. We found that growth temperature and the flow rate of hydrogen are two key factors influence the vibrational properties of CVD-grown $MoSe_2$. In sample grown at higher growth temperature and with higher hydrogen flow rate, a new peak emerges at ~250 $cm^{-1}$ and strengthens, which demonstrates that Se vacancies can be introduced by thermal annealing and hydrogen etching in CVD processing. DFT calculation results are in good agreement with the experiment data and indicate that the defect-related mode at 250 $cm^{-1}$ is an $A_{1g}$-like localized mode bounded to $V_{Se}$, which is confirmed by polarized Raman scattering experiments. Our findings solve a long-standing problem in literature and invalidate recent attempts to assign this peak in 2D MoSe$_2$[27, 34, 36-38]. And our methods can be extended to other kind of 2D TMDCs or TMDCs alloys to study their vibrational properties.

## 2. Methods

**Sample preparations:** The CVD processes to synthesize MoSe$_2$ nanosheets is like our previous work to grown WSe$_2$[39]. To tailor the Se vacancies, the growth temperature and the flow rate of hydrogen were tuned. The total flow rate of Ar and H$_2$ was kept at the fixed value of 58 sccm.

**Raman scattering measurements:** The normal Raman spectra and mappings were performed in a Renishaw inVia Raman microscope equipped with a 532 nm laser. Polarized Raman scattering measurements were performed in a NT-MDT NTEGRA Spectra Raman system with 532 nm excitation (See Figure S6).

**Density functional theory (DFT) calculations:** Vibrational and Raman properties were calculated by Vienna ab initio simulation package (VASP) using the projector-augmented wave (PAW) methods and Perdew-Burke-Ernzerhof (PBE) functional with D3 correction (DFT-D3). Three different V$_{Se}$ concentrations (pristine, 11.1% and 25%) were concerned (see details in Supporting Information).

## 2. Results and discussions

### 2.1 Heating effect and hydrogen etching in CVD process

MoSe$_2$ monolayers are synthesized by CVD method as most works utilized, with MoO$_3$ and Se powders being the precursors (see details in Experimental section). Figure 1(a) shows the optical image of a MoSe$_2$ monolayer with a second layer at the central area, identified by the PL spectra (insert). Raman spectra and Raman intensity mapping image at 351 cm$^{-1}$ (B$^1_{2g}$ mode in multilayer) in Figure 1(b) also confirm it. Except for the A$_{1g}$ mode at ~239 cm$^{-1}$ and E$^1_{2g}$ mode at ~288 cm$^{-1}$, a remarkable peak located at about 250 cm$^{-1}$ is also commonly found in our CVD grown MoSe$_2$ nanosheets. This peak has been proved to be a Se vacancy defect-activated mode by Masoud et al[37]. In their work, the losses of Se in the MoSe$_2$ powder precursors are mainly caused by the laser heating during the first 2 minutes of their laser-based synthesis process. We conjecture the heat etching effect also exists in CVD procedures. In another hand, a small amount of hydrogen is always introduced in reducing MoO$_3$ reaction. The hydrogen may react with the just grown MoSe$_2$ nanosheets during growth and cooling procedure. Similar effect of hydrogen etching can be traced in CVD processing of graphene growth[40].

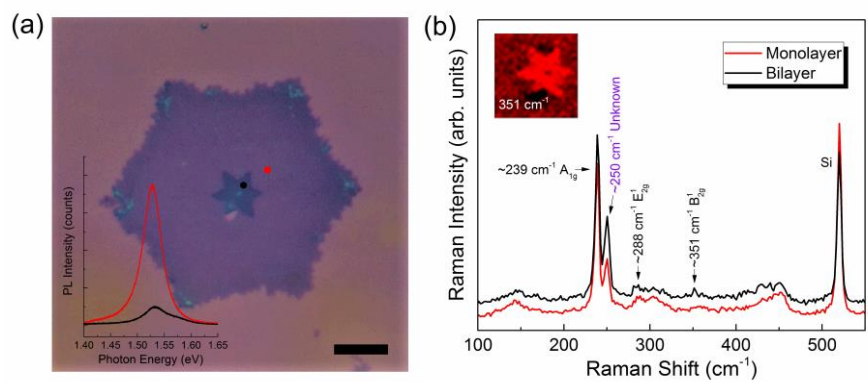

Figure 1 (a) Optical image and PL spectra of as prepared a MoSe$_2$ nanosheet. Scale bar is 10 μm. (b) Raman spectra of the monolayer and the bilayer MoSe$_2$. Insert is the Raman mapping image at the bilayer region

So we designed experiments using four different parameters. The correlated Raman results are illustrated in Figure 2. With higher growth temperature and higher flow rate of hydrogen, the relative intensity of peak D (defect-related peak around 250 cm$^{-1}$) to peak $A_{1g}$ increases. Besides, peak $A_{1g}$ weakens, shifts from 239.4 cm$^{-1}$ to 234.1 cm$^{-1}$, and D peak has a slightly blue shift from 250 cm$^{-1}$ to 253 cm$^{-1}$. These results show a good agreement with the experimental results in Masoud et al.'s work[37]. So we successfully tailored the Se vacancy concentrations by CVD method, and found that growth temperature and the flow rate of hydrogen are two key factors inducing Se vacancies in CVD process of synthesizing MoSe$_2$ nanosheets. In addition, the etching effects of heating and hydrogen can also be confirmed by the Raman mapping images of a MoSe$_2$ monolayer shown in Figure 3.

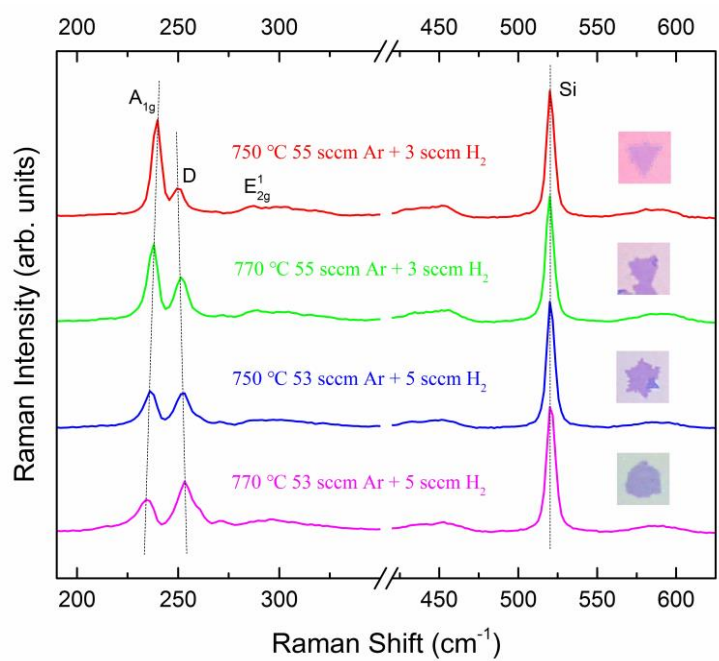

Figure 2 Raman spectra of MoSe$_2$ monolayers prepared by CVD at different growth parameters. The notations of the Raman peaks used here are same with that in Figure 1 for simplicity.

Figure 3(a) and Figure 3(b) exhibit the Raman intensity mapping images of $A_{1g}$ and D peaks. Good homogeneity can be seen all over the monolayer except for the edge area, since it cannot be totally irradiated by laser spot (~1 μm diameter). Figure 3(c) and 3(d) show the variation of Raman intensity ratio between of $A_{1g}$ and D modes. From center to edge, the ratio gradually turns smaller, i.e. the central areas of the nanosheets have larger amount of defects. In WS$_2$ CVD synthesis processes reported in some previous works, similar etching mechanism was also discovered, that is the photoluminescence intensity gradually changes from center region to edge of the triangle WS$_2$ single layer[41]. It means that etching is concomitant with the growth of the crystal. Furthermore, it may also happen at the cooling procedure. As seen in atomic force microscope (AFM) images in Figure S1, the MoSe$_2$ nanosheet degrades about 800 nm and some nanoparticles regrow around the edges, as the same reversible reaction mechanism during cooling stage as reported by Bo Li et al[42]. Meanwhile, some vacant triangular or rhombic areas induced by etching can be seen clearly in the AFM phase image. So in order to avoid these etching effects in CVD processing, we used less hydrogen flow rate (2 sccm) and opened the furnace immediately after growth for fast cooling, then the D peak almost disappeared as shown in Figure S2.

Ismail et al. attributed the disappearing of D peak (assigned to 2ZA(M) in their work) to the degradation of the high-quality crystal after aging 6 months[36]. However, in our work, due to the Se vacancies, the samples are quite unstable even under a vacuum environment, according to the concentration of Se vacancies. The samples with high vacancy concentration are damaged after two month's aging, while the samples with low vacancy concentration remain unchanged (see Figure S3). Recently, when preparing our manuscript, the similar etching effect of hydrogen was also discovered in $MoS_2$ grown by chemical vapor deposition for photoluminescence enhancement[43]. It indicates that this approach can be extended to all other TMDCs for a wide range of physical property studies.

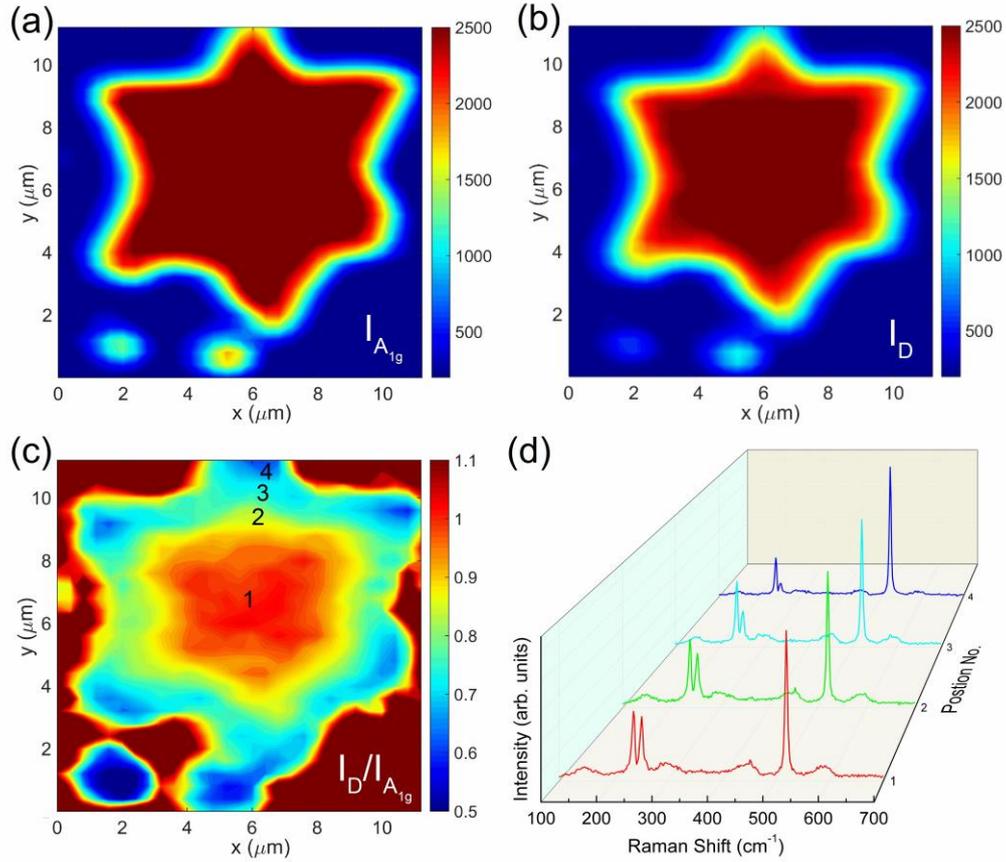

Figure 3 (a) Raman intensity mapping image of $A_{1g}$ mode. (b) Raman intensity mapping image of D mode. (c) The intensity ratio mapping image of D mode to $A_{1g}$ mode. (d) Four Raman spectra at different locations of the $MoSe_2$ monolayer as denoted in (c).

## 2.2 Behaviours of $A_{1g}$ and $E^1_{2g}$ modes

It can be seen in Figure 2 that as Se vacancy concentration increases, except the down shifting, $A_{1g}$ mode also shows an asymmetric broadening to lower wavenumber, while $E^1_{2g}$ mode seems unchanged. Previously, slight up shift of $A_{1g}$ mode and down shift of $E^1_{2g}$ mode were observed in $MoS_2$[43], but both $A_{1g}$ mode and $E^1_{2g}$ mode in $WS_2$ show down shift[44]. We found that their behaviors with increasing defect concentrations follow the trend of their phonon dispersion when the wave vector gets away from Brullioun zone center point. It means that the phonons around Brullioun zone center or at Brullioun zone edge participate in the Ramam process. A phonon confinement model can well explain the evolution of these two first-order optical phonon modes [45]. The phonon dispersion of monolayer $MoSe_2$ is shown in Figure S5, the two branches of $E^1_{2g}$ mode, LO and TO phonon bands are relatively flat throughout the Brullioun zone, which agrees well with the behavior of $E^1_{2g}$ mode in Raman spectra. The $A_{1g}$ phonon band shows obvious down shift along Γ–K and Γ–M. We calculated the line shapes of $A_{1g}$ mode at different vacancy concentrations using the RWL model developed by Tan et al[44]. The Se vacancy concentrations of all samples are characterized by X-ray photoelectron spectroscopy (see Figure S7).

Simulation results are presented in Figure 4, $L_D$ means the grain size or the average distance between defects. As we can see, the line shapes and peak frequency shifts of $A_{1g}$ mode are well fitted at all Se vacancy concentrations, except the additional Se vacancy-induced peak at ~250 cm$^{-1}$. In addition, at 8% and 14% Se vacancy concentrations, there is another peak at around 215 cm$^{-1}$. According to the phonon dispersion, 215 cm$^{-1}$ is just corresponding to the frequency of $A_{1g}(M)$, so it can be assigned to the $A_{1g}(M)$ due to the phonon confinement effect.

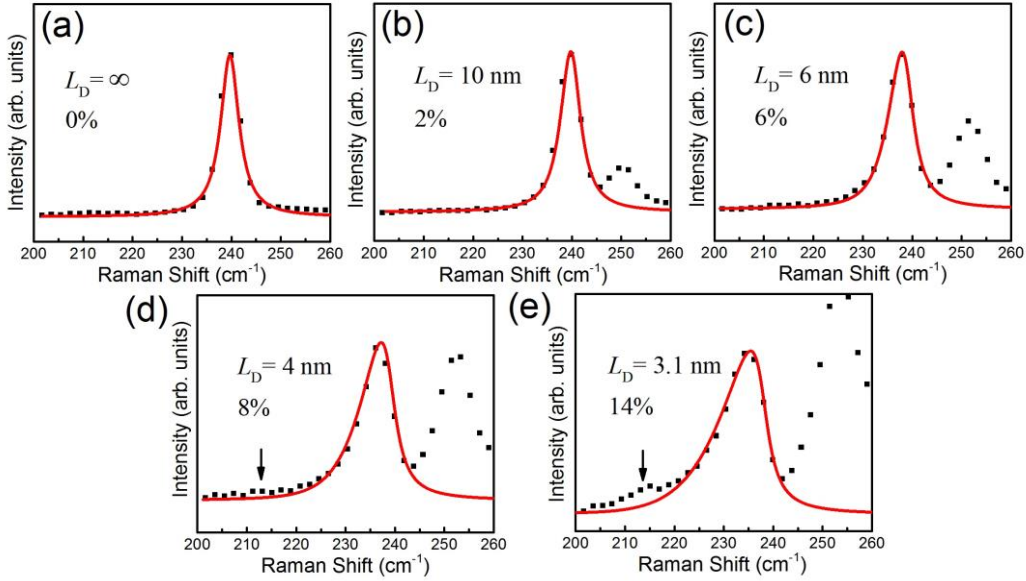

Figure 4 Simulation results of Raman spectra at (a) 0%, (b) 2%, (c) 6%, (d) 8% and (e) 14% Se vacancy concentrations. The corresponding fitting parameters $L_D$ are (a) ∞, (b) 10 nm, (c) 6 nm, (d) 4 nm and (e) 3.1 nm. All fitting parameters α is 0.008. Red line represents the calculated Raman curve and black symbol represents the experimental data.

## 2.3 DFT calculation of $A_{1g}$ and D modes

So far we have demonstrated that growth temperature and flow rate of hydrogen are two key factors that can introduce and tailor Se vacancies in CVD grown MoSe$_2$ nanosheets. Besides, simulation results based on phonon confinement model give a perfect explanation on behaviors of intrinsic peaks of monolayer MoSe$_2$ with increasing Se vacancy concentrations. However it can not predict the new defect-activated peak in Raman spectra. In Masoud et al.'s work, both single Se vacancy ($V_{Se}$), and the column Se vacancy ($V_{Se}$) were detected by Z-contrast STEM[37]. However, they failed to have a deep understanding of the single Se vacancy-activated Raman peak observed in experiments using density functional theory (DFT) calculation. No peak appeared at ~250 cm$^{-1}$ in their calculated Raman spectra. So in order to confirm this Se vacancy defect type and to give a better understanding of this defect-related mode, we also utilized DFT simulations and three different $V_{Se}$ concentrations were considered, i.e. 0% (pristine), 11.1% and 25% (see details in Surporting Information). With a Se atom removed from the lattice, the symmetry of MoSe$_2$ film becomes lower and more vibrational modes appear at the Brillouin zone center. Among all of these complicated vibrational modes, we can easily distinguish the $A_{1g}$ mode from others by analyzing the vibrational images of the Raman modes.

Figure 5 shows the vibrational images of $A_{1g}$ mode at different vacancy concentrations. The atomic vibration of $A_{1g}$ mode in pure, perfect MoSe$_2$ monolayer is that Mo atomic layer is stationary while the bottom and top layers of Se atoms vibrate in opposite directions along z axis. Briefly, in defective samples the general behavior of the two layers of Se atoms remain unchanged, except that the vibration directions of some Se atoms show different degrees of deviation to *z* axis, and Mo atoms participate in the vibration. This can be seen more clearly in the animation of this vibrational mode ($A_{1g}$ 11.1%.mp4).

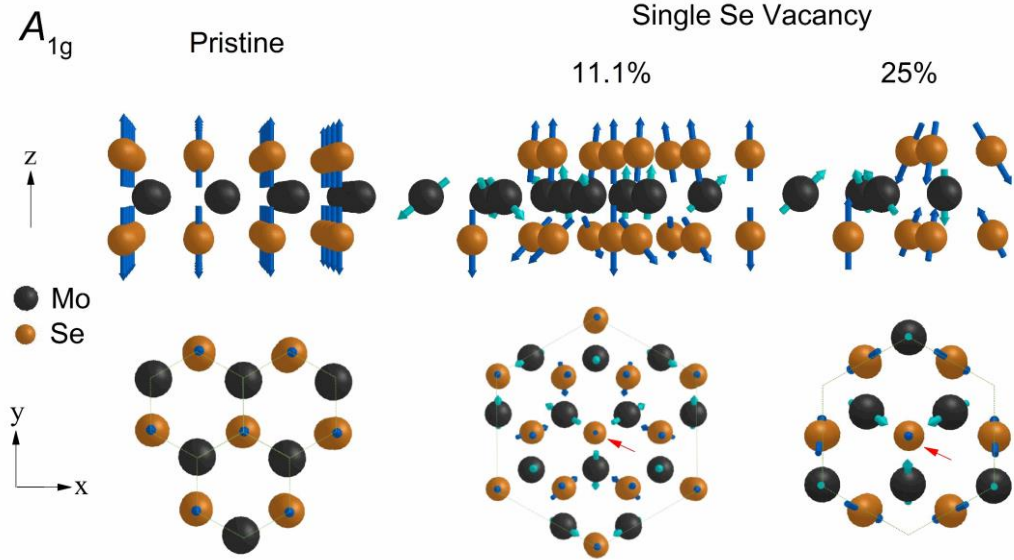

Figure 5 Vibrational images of $A_{1g}$ mode at different $V_{Se}$ concentrations. The upper line are the side views along z axis, the bottom line are the top views. The green hexagonal line means one unit cell. The red arrows point to the locations of $V_{Se}$.

In detail, after removing a top Se atom from the lattice, the unit cell remains to be hexagonal but centered by the Se vacancy. The Se atom under the $V_{Se}$ still vibrates along *z* axis. The motions of the six nearest neighbor bottom Se atoms are affected most in 11.1% vacancy sample. At a further distance, the twelve second nearest neighbor Se atoms are not affected. As in the 25% vacancy sample, all neighbor Se atoms are influenced by the Se vacancy because the distances between Se vacancies are shorter at higher vacancy concentrations. These variations of motions of Mo atoms and Se atoms can be directly related to red shift of $A_{1g}$ mode. As shown in the calculated Raman spectra in Figure 6(a), the frequency of $A_{1g}$ mode is decreased by 16 cm$^{-1}$ from 0% to 25%. The calculated phonon frequecy shift of $A_{1g}$ shows highly agreement of previous work[37].

Now we focus on the defect related mode. As seen in Figure 6(a), except for the red-shifted $A_{1g}$ mode, a defect induced peak emerges at ~254 cm$^{-1}$ at 11.1% vacancy concentration and it shifts to 260 cm$^{-1}$ when the concentration increases to 25%. We assign it to the D mode. The calculated Raman spectra show great agreement with our experimental results, except the strength of D mode is underestimated. We extract the vibrational images and Raman tensors of D mode. As shown in Figure 6(b), D mode is also an out-of-plane mode and the vibrational behavior of D mode is quite similar with $A_{1g}$ mode in defective samples. The main difference between D mode and $A_{1g}$ mode is that the vibrational direction of the Se atom under the vacancy is opposite with other Se atoms in the bottom layer. Another distinct feature is that, at 11.1% $V_{Se}$ concentration, the three nearest Mo atoms of $V_{Se}$ and the Se atoms have much larger vibrational amplitude. At the six corners of unit cell, the amplitude of bottom Se atoms becomes smaller and the six top Se atoms are static due to longer distance from the $V_{Se}$ (see more clearly in D 11.1%. mp4). This is a typical picture of local mode [46], indicating that D mode may be the localized $A_{1g}$ mode bounded to $V_{Se}$. Surprisingly, the D mode has the same form of Raman tensor with $A_{1g}$ mode [47]. Our calculations can well explain the correlation between D mode and $A_{1g}$ mode discovered by previous works[34, 36].

To further confirm this, polarized Raman spectra of samples at three different Se vacancy concentrations are collected. As shown in Figure 6(c), at cross configuration, the $A_{1g}$ peak nearly disappears while $E^1_{2g}$ peak remains unchanged which highly consist with theoretical analysis and experimental results in other works[48, 49]. The D peak also disappears at cross configuration. Additionally, Figure 6(d) illustrates the polar plots of Raman intensity as function of the polarized direction of incident light. As expected, the line shape of $A_{1g}$ mode shows a typical dumbbell structure as previous works found[48]. D mode has the same structure which agrees very well with our calculation results.

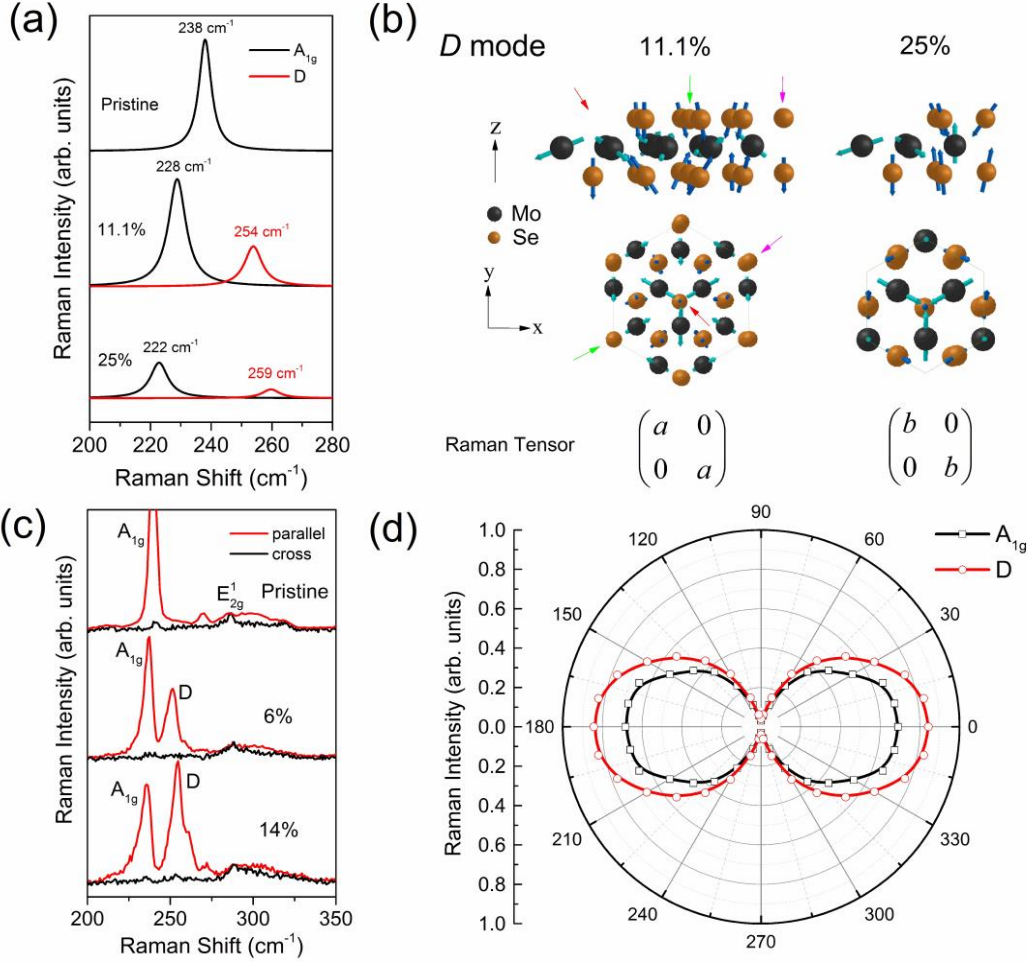

Figure 6 Calculated Raman spectra of MoSe$_2$ at different VSe concentrations. (b) The vibrational images and Raman tensor of D modes at 11.1% and 25% $V_{Se}$. Red arrows point to the $V_{Se}$, green and magenta arrows point to the two unequivalent sites at corners of the hexangnal unit cell. (c) Polarized Raman spectra of three MoSe$_2$ sample with 0%, 6% and 14% VSe concentrations. (d) Polar plot of the Raman intensity as a function of $\phi$ for $A_{1g}$ and D modes.

## 3. Conclusions

In summary, using CVD methods we have successfully tailored the concentrations of Se vacancies in two-dimensional MoSe$_2$ through tuning the growth temperature and flow rate of hydrogen. The characteristic $A_{1g}$ peak of MoSe$_2$ in the Raman spectra shifts to lower energy and a new peak (labelled as D) at ~250 cm$^{-1}$ emerges and shows a little blue shift with increasing Se vacancy concentrations. DFT calculations demonstrated that D peak originates from single Se vacancy and it is an $A_{1g}$-like out-of-plane vibrational mode localized to the single Se vacancy. The polarized Raman spectra confirmed our calculation results. Our findings may pave a way to tailor the chalcogenide element vacancies in two-dimensional TMDCs by CVD method, and to investigate their vibrational properties with defects.

**Supporting information**

Supporting Information is available from the Wiley Online Library or from the author.

**Acknowledgements**

This research was financially supported by the National Natural Science Foundation of China (No. 11304060) and the Foundation of Harbin Institute of Technology for the Incubation Program of the development of basic research outstanding talents (No. 01509321) and the China Postdoctoral Science Foundation.

# Supporting Information

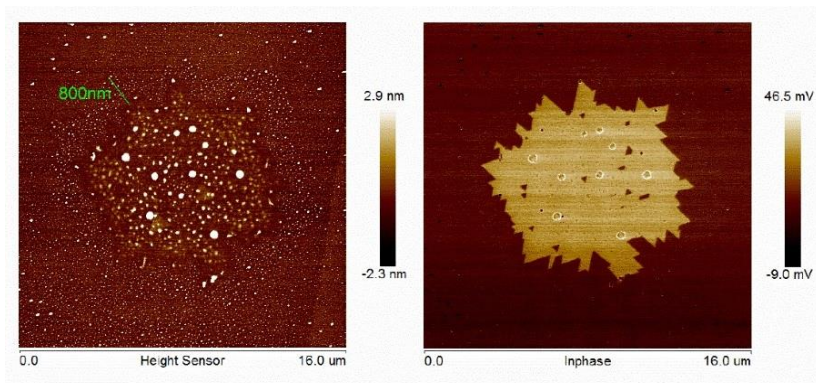

**Figure S1.** AFM height (left) and phase (right) images of a MoSe$_2$ monolayer.

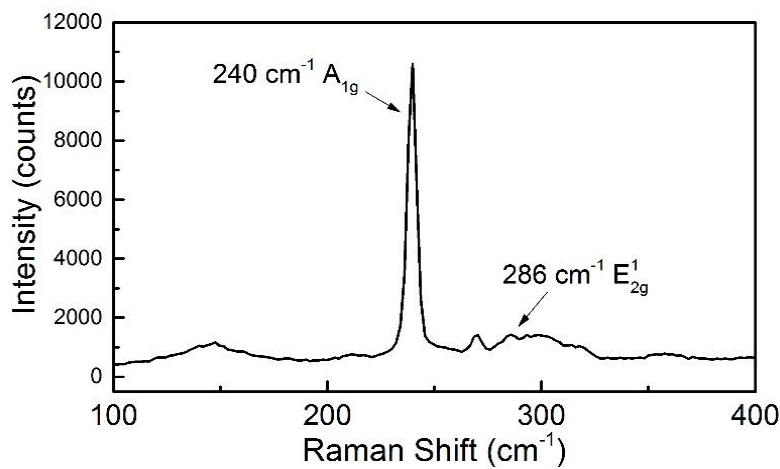

**Figure S2** Raman spectrum of monolayer MoSe$_2$ without Se vacancies

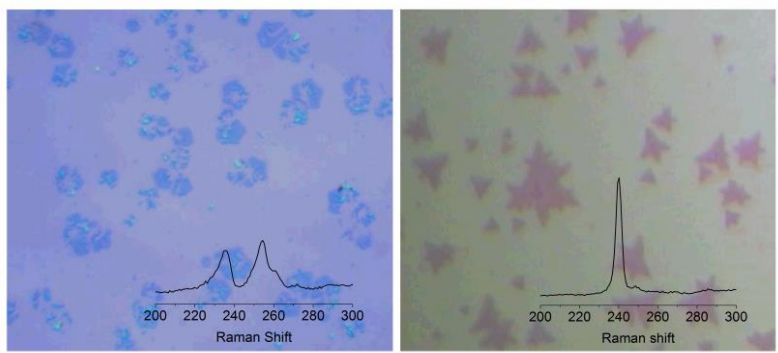

**Figure S3** Optical images of MoSe$_2$ nanosheets at two different Se vacancy concentrations after two months aging. Inserts are the corresponding Raman spectra which can indicate the quality of MoSe$_2$ nanosheets. Samples in left picture were damaged by air while samples in right remain perfect after two months aging.

**DFT calculations for vibrational and Raman properties**

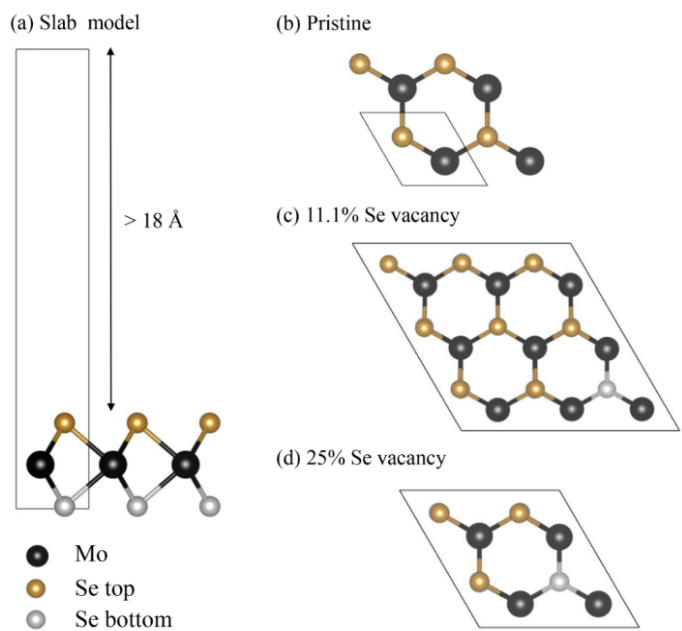

**Figure S4.** (a) The slab model of 3×3 monolayer MoSe$_2$. The slab model is extended from the primitive cell. The top views are of (b) pristine, (c) 11.1% V$_{Se}$, and (d) 25% V$_{Se}$ structures.

To understand the Raman spectrum of detective MoSe$_2$ films, first-principle density functional theory (DFT) calculation of MoSe$_2$ monolayer models containing different Se defect concentrations were carried out. Vacuum spacing in the out-of-plane direction is set to be more than 18 Å to avoid spurious interaction. Structure and phonon properties are calculated by Vienna ab initio simulation package (VASP) using the projector-augmented wave (PAW) methods and Perdew-Burke-Ernzerhof (PBE) functional with D3 correction (DFT-D3). The energy cutoff was set as 400 eV and residual forces for ion iteration as 10$^{-4}$ eV/Å.

The vacancy concentration is defined as the number of Se vacancy sites divided by the total number of Se atoms in the defective Se layer. We have considered models of three Se defect concentrations: pristine, 11.1% (3 × 3 supercell with a single Se vacancy) and 25% (2 × 2 supercell with a single Se vacancy), as shown in Figure S3. For a primitive unit cell of pristine MoSe$_2$, it was optimized with Γ-centered 24×24×1 Monkhorst-Pack (MP) k-points and the optimized in-plane lattice constant is 3.29 Å. The Γ-centered MP grids were set as 12×12×1 and 16×16×1 for 11.1% detective model and 25% detective model respectively.

After the structural relaxations, the normal vibration modes are calculated within frozen phonon approximation. Giving these vibration mode, Raman tensors and intensities are obtained based on the calculation of the derivatives of the dielectric tensors with respect to phonon vibration using the script "vasp_raman.py" [1]. This is a Raman off-resonant activity calculator using VASP as a back end. Lorentz broadening of the modes' intensities yielded the Raman spectrum.

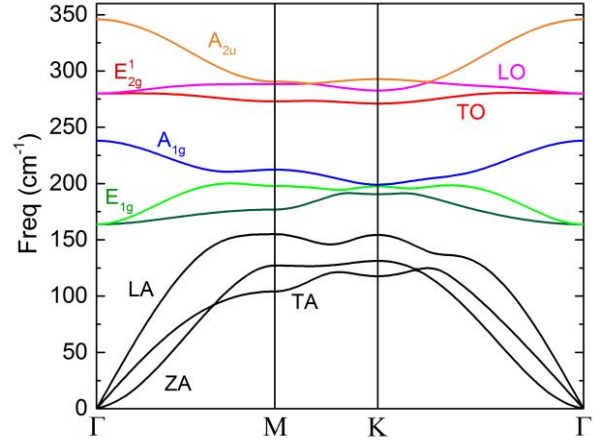

**Figure S5** Calculated Phonon dispersion of monolayer MoSe$_2$

**Polarized Raman scattering**

As shown in Figure S6 the polarization direction of scattered light's electric field is fixed in an arbitrary angle, $\theta$, to x axis. The induced angle, $\phi$, between the polarization directions of incident light and scattered light is tuned from 0° to 360°. The Raman intensity for a given configuration can be determined by Raman tensor[2]:

$$I \propto \left| \hat{g}_s \cdot \tilde{R} \cdot \hat{g}_i \right|^2$$

where $\tilde{R}$ is the Raman tensor of the Raman mode, $g_s = (\cos\theta, \sin\theta)$, $\hat{g}_i = (\cos(\theta+\phi), \sin(\theta+\phi))$. According to the Raman tensors given in the main article, the intensities of A$_{1g}$ mode and D mode are both proportional to $\cos^2 \phi$, consistent with our experimental results shown in Figure 6d. Cross and parallel configurations mean that $\phi$ is 90° and 0° respectively.

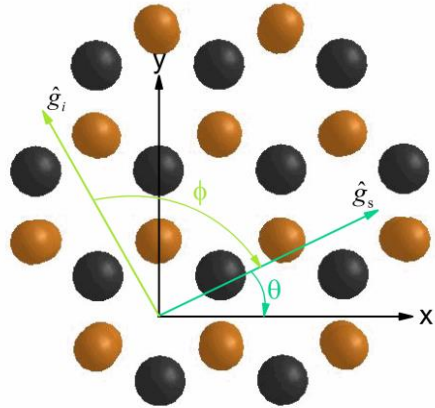

**Figure S6.** The diagram of Polarized Raman scattering experiments in Cartesian coordinate systems. $\hat{g}_s$ is the direction of scattered light and $\hat{g}_i$ is the direction of incident light. *x*- and *y*-directions represent zigzag and armchair directions respectively.

**X-ray photoelectron spectroscopy analysis**

The X-ray photoelectron spectroscopy (XPS) was adopted to determine the Se/Mo ratio in MoSe$_2$ samples with different Se vacancy concentrations. As seen in Figure S7, the doublet of Mo$^{4+}$ core levels are located at 229.1 eV (3d$_{5/2}$) and 232.2 eV (3d$_{3/2}$), and doublet of Se$^{2-}$ 3d core levels are located at 54.4 eV (3d$_{5/2}$) and 55.4 eV (3d$_{3/2}$). The difference value (~174.7 eV) between the 3d$_{5/2}$ peaks of Mo$^{4+}$ and Se$^{2-}$ confirms the Mo-Se bond.[3] It should be noted that the 3d peak of Mo$^{6+}$ and 3s peak of Se$^{2-}$ have an overlap with 3d peak of Mo$^{4+}$.[4] MoO$_3$ nanoparticles arising from the raw material should be responsible for the Mo$^{6+}$ peak in XPS, which can be identified by AFM image in Figure S1. Se/Mo ratios are estimated by the ratios of peak areas of Mo$^{4+}$ (3d) and Se$^{2-}$ (3d) concerning the element sensitivity factors (1.6 for Se, 11.008 for Mo).[5]

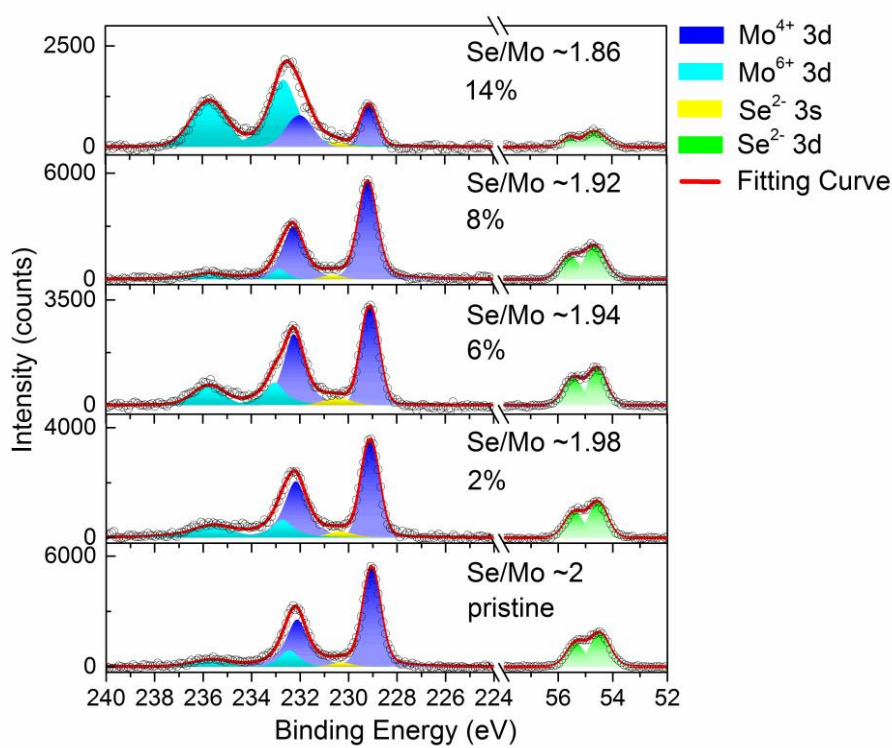

**Figure S7.** XPS spectra of MoSe$_2$ samples with different Se vacancy concentrations.